\documentclass[a4paper]{panl}
\usepackage{cite}
\usepackage{wrapfig}
\usepackage{graphicx}
\usepackage{amssymb}
\usepackage{amsfonts}
\usepackage{amsmath}
\usepackage{longtable}
\usepackage{rotating}
\usepackage{lscape}
\usepackage{epsfig}
\usepackage{multirow}
\usepackage{hyperref}

\usepackage{lineno}

\originalTeX
\begin{document}
\modulolinenumbers[1]

\issuearea{Physics of elementary particles and atomic nuclei. Experiment}

\title{
Central diffraction and ultra-peripheral collisions\\ in ALICE in Run 3 and 4}
\maketitle
\authors{
N.\,Burmasov\,$^a$\footnote{E-mail: nazar.burmasov@cern.ch}, 
for the ALICE Collaboration
}
\from{$^{a}$\,Petersburg Nuclear Physics Institute named by B.P.Konstantinov of NRC "Kurchatov Institute", 188300, 1 mkr. Orlova roshcha, Gatchina, Russia}

\begin{abstract}
В настоящее время детекторные системы эксперимента ALICE на Большом Адронном колайдере проходят крупную модернизацию в течение Long Shutdown~2 (2019-2021). В частности, время-проекционная камера (TPC) будет оборудована считывающими камерами на основе газовых электронных умножителей, а считывающая электроника некоторых других детекторов будет заменена на более быструю и гибкую, что позволит  \break ALICE собирать данные с большинства детекторов в непрерывном режиме и записывать Pb--Pb события в течение Run~3 (2021-2024) и Run~4 (2027-2030) при частотах, приблизительно равных 50~кГц. Коллаборация ALICE также рассматривает возможность сбора большого количества данных протон-протонных столкновений при частотах взаимодействий порядка 1~МГц с использованием предварительного отбора редких событий в режиме онлайн и оффлайн. Для выполнения этих задач требуется разработка совершенно новой вычислительной системы, которая будет использоваться для быстрой реконструкции событий и сжатия потока данных. В этих условиях разработка стратегии отбора становится нетривиальной задачей для случая редких событий, таких как дифракционные в протон--протонных столкновениях и ультра-периферические в ядро-ядерных свинцовых, которые характеризуются отсутствием частиц в передней и задней области быстрот и наличием лишь нескольких треков в центральной области. В данной работе приведены мотивация изучения центральных дифракционных и ультра-периферических событий, а также исследование возможности отбора таких событий в Run~3 и Run~4. 

\vspace{0.2cm}

The ALICE experiment at the LHC is undergoing a major upgrade during the Long Shutdown~2~(2019-2021). In particular, the Time Projection Chamber (TPC) is being equipped with new GEM-based readout chambers and the readout electronics of several detectors are being replaced with faster and more flexible technology. This will allow ALICE to read out most of the detectors in the continuous mode and record minimum bias Pb--Pb events at rates of about 50~kHz in Run~3 (2021-2024) and Run~4 (2027-2030). The ALICE Collaboration is also considering the possibility to collect a large sample of proton--proton collisions at interaction rates of about 1~MHz using online and offline preselection of rare events. These goals require a completely new online computing system that will be used to perform fast reconstruction and compression of the data stream. The event selection strategy becomes especially challenging for the case of central diffractive events and ultra-peripheral Pb--Pb collisions characterized by rapidity gaps at forward and backward directions with only few tracks at central rapidity. In this contribution, the motivation for studying central diffractive and ultra-peripheral events is presented, and feasibility studies for their selection in Run~3 and 4 will be given.
\end{abstract}
\vspace*{6pt}


\section{Introduction}

ALICE (A Large Ion Collider Experiment)~\cite{Aamodt:2008zz} is one of the four major experiments at the Large Hadron Collider, CERN. It is dedicated to the study of heavy-ion collisions at extremely high energies and probing the characteristics of the quark-gluon plasma. The detector is designed to cope with high particle densities in Pb--Pb collisions at the LHC: it has high detector granularity, a high tracking efficiency down to transverse momenta of $p_{\rm T} \simeq$ 0.15~GeV/$c$ and excellent PID capabilities up to 20~GeV/$c$. 

During the Long Shutdown~2 the ALICE detector systems are being upgraded significantly. In particular, the Time Projection Chamber (TPC) is being equipped with new GEM-based readout chambers and the readout electronics of several detectors is being replaced with faster and more flexible technology. The upgrade will allow ALICE to replace the triggered readout of Run 1 and Run 2 by a continuous readout of all events and switch to a full online event processing.  The new readout regime will make possible the recording of minimum bias Pb--Pb events at rates of about 50~kHz and proton--proton collisions at interaction rates of about 1~MHz in Run~3 (2021-2024) and Run~4 (2027-2030) leading to an increase of the data sample by a factor of 100~\cite{Buncic:2015ari}. The computing scheme will also be replaced with the new $\textup{O}^2$ computing framework~\cite{Eulisse:2019irt}. Data processing will be divided in two stages: the synchronous (or online) and asynchronous (or offline). The synchronous processing is needed for detector calibration and data compression. Synchronous data processing will result in Compressed Time Frames (CTF) -- compressed chunks of data, corresponding to events that are recorded in data chunks covering $\sim$ 10~ms. The CTFs and the calibration data will be stored to an on-site disk storage, and then copied to tapes. When the $\textup{O}^2$~computing farm is not intensively used for the synchronous processing, it will carry out a part of the offline reconstruction, which reprocesses the data and generates the final reconstruction output. The part of offline processing that exceeds the capabilities of the on-site farm will be done in the LHC Computing Grid~\cite{Rohr:2020qct}.

In the following sections the motivation for the studies of central diffractive pp events and ultra-peripheral Pb--Pb collisions in Run 3 and Run 4 is discussed. Particularly, the topics of interest for central diffraction research are related to glueball searches, pomeron spin structure and rare resonance studies~\cite{future_pp_programme}. Ultra-peripheral collisions in ALICE allow for more precise measurements of poorly known gluon distribution functions in nuclei at small Bjorken-$x$ and could be possibly used for probing QED and BSM physics via light-by-light (LbyL) scattering process and axion-like particle (ALP) searches~\cite{Citron:2018lsq, atlas_lbyl, Sirunyan_2019}.

After the upgrade, the ALICE detector will be able to read out Pb--Pb events at rates of about 50~kHz and pp events at rates of the order of 1~MHz. In view of the increasing data stream from the detector and the foreseen continuous readout mode, the event selection strategy needs to be adjusted accordingly. This task becomes especially challenging for the case of rare central diffractive events and ultra-peripheral Pb--Pb collisions characterized by rapidity gaps at forward and backward directions with only few tracks at central rapidity. In this contribution, possible event selection strategies for Run~3 and Run~4 are considered and the results of feasibility studies utilizing Monte~Carlo event generators and the $\textup{O}^2$~framework for full detector system simulations are shown.
\clearpage

\section{Central diffraction in proton--proton collisions}

Central diffractive pp collisions at the LHC may be used for investigations of the strong interaction in the non-perturbative regime. Central exclusive production (CEP) at low masses is usually described in terms
of a double pomeron exchange mechanism~\cite{forshaw_ross_1997} resulting in enhanced production of meson resonances with $I^G(J^{PC}) = 0^+ (0^{++} , 2^{++} , ...)$ quantum numbers and gluonic bound states. According to lattice QCD calculations, the lightest glueballs are expected to have masses of $M^G (0^{++})$ = 1710~MeV and $M^G (2^{++})$ = 2390~MeV while experimental measurements suggest $f_0 (1370)$, $f_0 (1500)$ and $f_0 (1710)$ mesons for the role of glueball candidates. The nature of these glueball-candidate states is still controversial and  higher-precision measurements at the LHC would provide additional important insight. For a review see e.\,g.~\cite{Crede_2009}.

CEP can also be used to investigate the spin structure of the pomeron and its coupling to hadrons. Though the pomeron was initially introduced as a Regge trajectory explaining the slow rise of hadronic cross sections at high energies and running over all allowed spin states, in the famous Donnachie-Landshoff model it has been phenomenologically described as an effective vector current~\cite{Donnachie:2002en}. On the other hand, gluon-ladder calculations and the observed $s$-channel helicity conservation in the photoproduction processes favor the scalar nature of the pomeron. However, the scalar pomeron is disfavored by recent measurements on the helicity structure of small-$t$ proton–proton high-energy elastic scattering from the STAR experiment~\cite{Adamczyk:2012kn}, while the tensor pomeron model~\cite{Ewerz:2013kda} was found to be consistent with the STAR data~\cite{Ewerz:2016onn}. Partial wave analysis of various final states produced in CEP events can be used as a tool to resolve this puzzle and study whether the pomeron effectively couples like a scalar, vector or tensor current.

\begin{figure}[h!]
\includegraphics[width=0.49\textwidth]{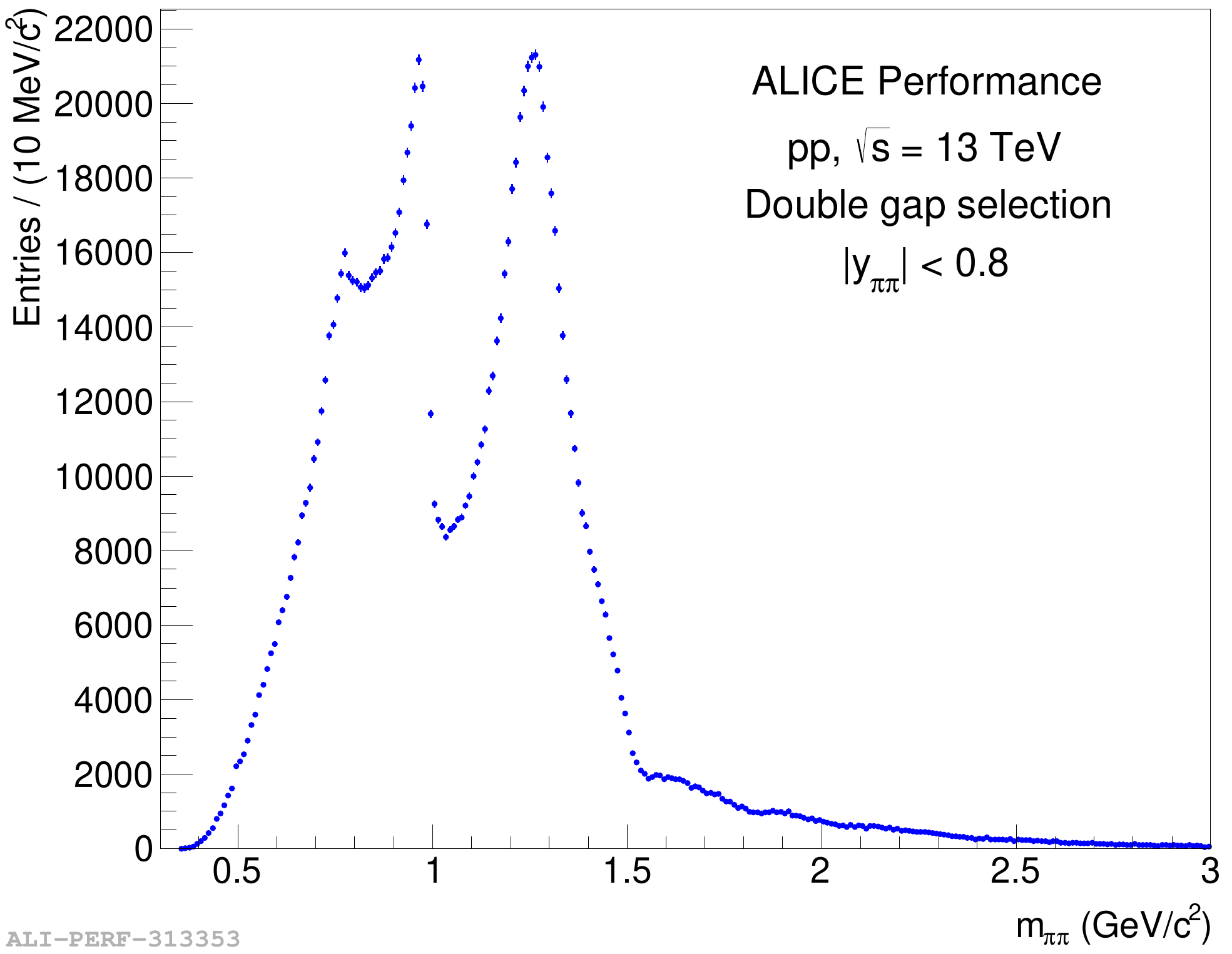}
\includegraphics[width=0.49\textwidth]{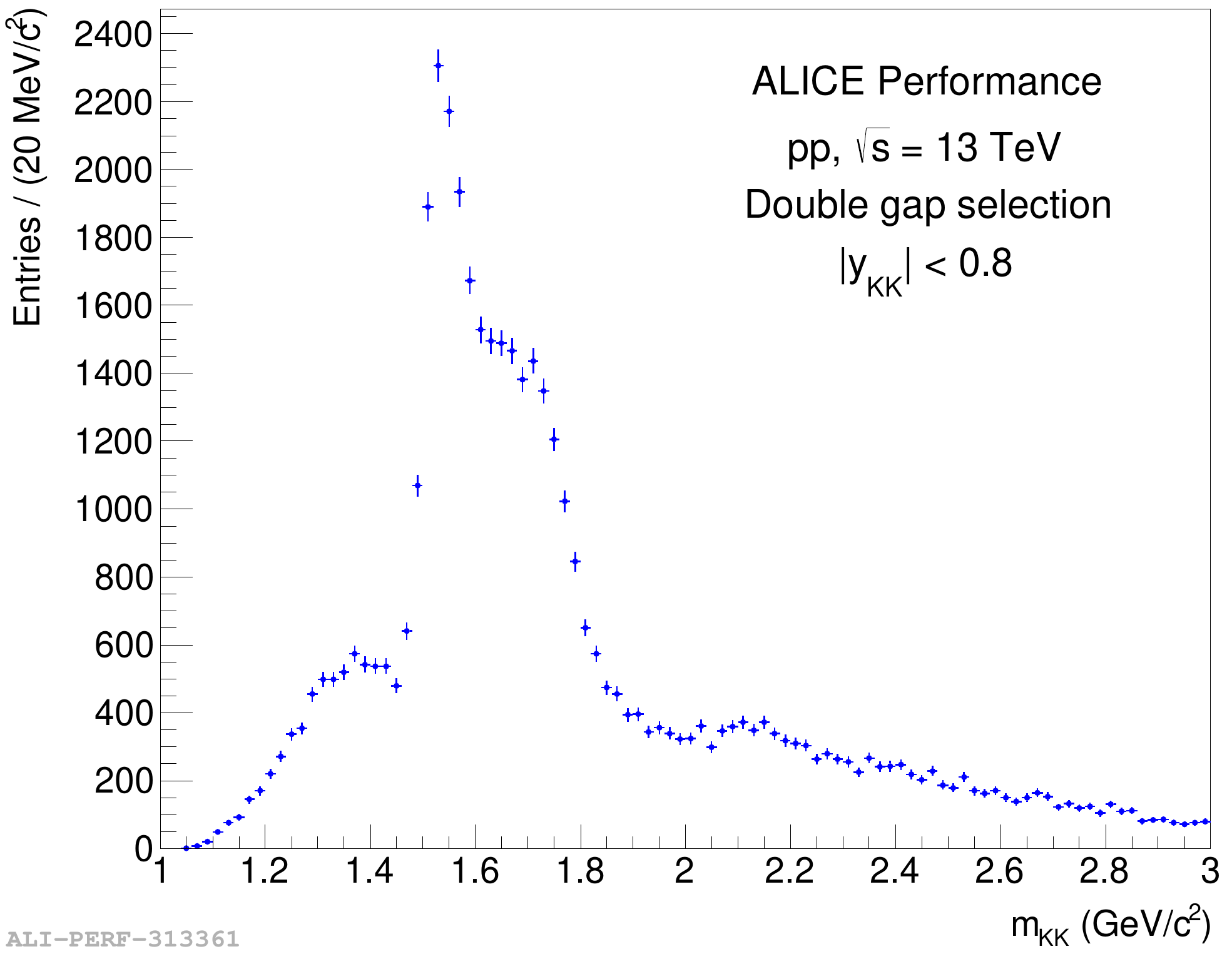}
\caption{Raw invariant mass spectra of pion (left) and kaon (right) pairs in central diffractive events collected by ALICE in proton--proton collisions at 13~TeV. From Ref.~\cite{Azzi:2019yne}.}
\label{fig:run2_data}
\end{figure}

During the LHC Run~2 a considerable amount of CEP data with integrated luminosity of about 8~$\textup{pb}^{-1}$ at $\sqrt{s}~$=~13~TeV have been collected (see Fig.~\ref{fig:run2_data}) and can already be used for analyses. The ALICE Collaboration plans to collect a significantly larger sample during Run~3~and~4 with an expected integrated luminosity of about 200~$\textup{pb}^{-1}$ at $\sqrt{s}=$~14~TeV profiting from much better efficiency in the continuous readout mode~\cite{future_pp_programme}.

\section{Ultra-peripheral heavy-ion collisions}

Relativistic heavy ions are sources of a strong electromagnetic field, which can be described in terms of equivalent quasi-real photon flux. In ultra-peripheral heavy-ion collisions (UPC) the impact parameter is larger than the sum of radii of the incoming particles, thus hadronic interactions are suppressed, photon-induced interactions being dominant. Consequently, UPC at the LHC are well suited for photon-photon and photonuclear interaction measurements.

At LO in perturbative QCD, the coherent photoproduction cross section is proportional to the square of the gluon density in nuclei, hence the photoproduction of vector mesons is particularly interesting. The ALICE experiment covers acceptance range which corresponds to a Bjorken-$x$ between $\sim$~$10^{−2}$ and $\sim$~$10^{−5}$, while the heavy-quark mass serves as a hard scale justifying perturbative calculations~\cite{Kryshen:2019ggx}. Thus, coherent photoproduction of charmonium in UPC can be used as a tool to probe poorly known gluon distribution functions and their modification in nuclei known as nuclear shadowing \cite{Guzey:2013xba}, which is an important factor in the description of the initial stages of heavy-ion collisions. 

The extraction of gluon shadowing from vector meson photoproduction measurements is hindered by the fact that the measured UPC cross section is expressed as a sum of two terms corresponding to low-$x$ and high-$x$ contributions since each of the colliding ions can serve as a photon source. The expected large data sample of UPC events in Run~3~and~4 may help to decouple low-$x$ and high-$x$ contributions in the cross section via measurements of rapidity-differential cross section with and without additional neutron activity in the Zero Degree Calorimeters~\cite{Guzey:2013jaa}, or by utilizing peripheral photoproduction cross sections~\cite{Contreras:2016pkc}.

\begin{figure}[h]
\centering
\includegraphics[width=0.65\textwidth]{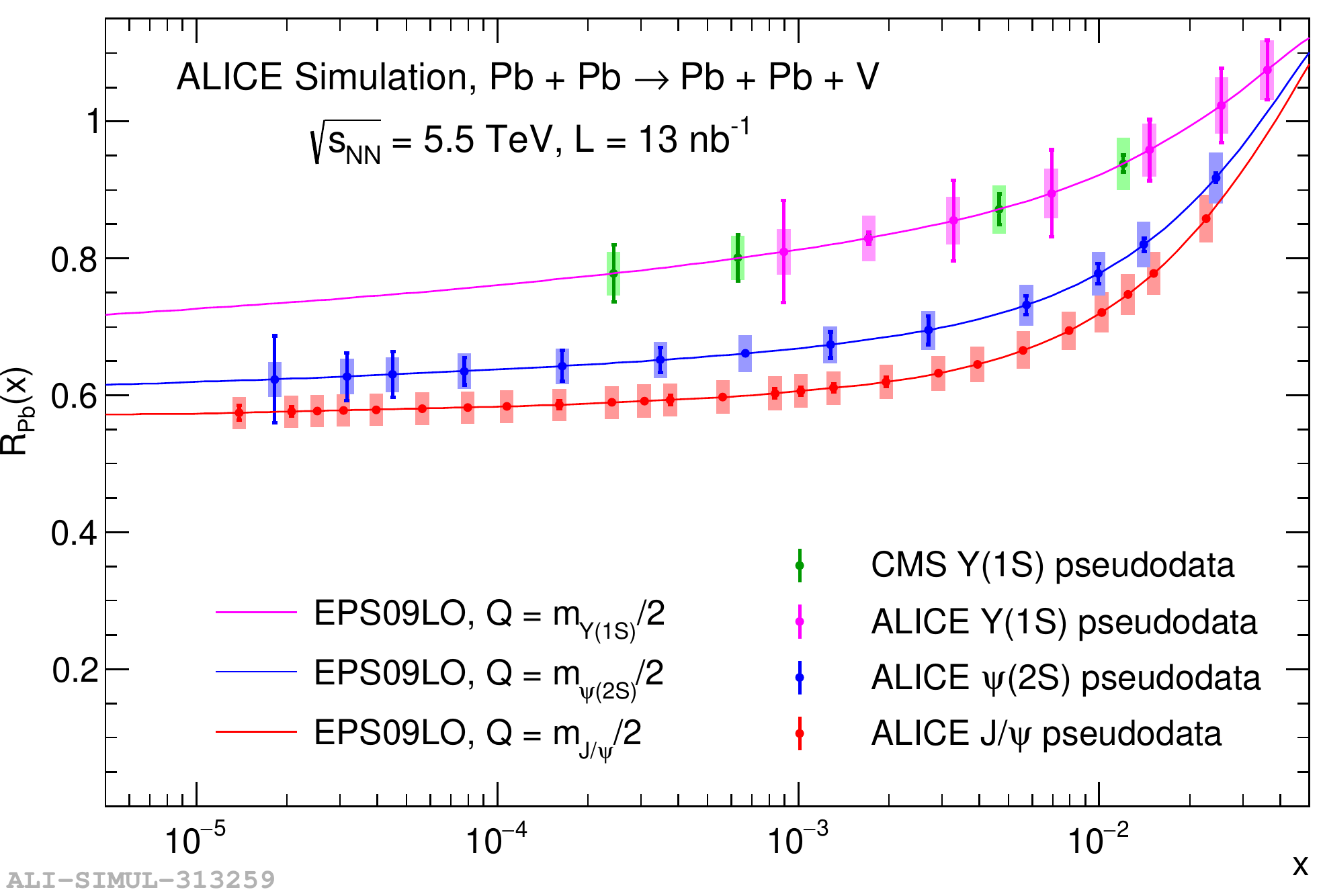}
\caption{Pseudodata projections for the nuclear suppression factor by ALICE and CMS measured with the photoproduction of three heavy vector mesons in Pb–Pb ultra-peripheral collisions. The pseudodata points are derived from EPS09-based photoproduction cross section projections. From Ref.~\cite{Citron:2018lsq}.}
\label{fig:rPb}
\end{figure}

The expected experimental uncertainties on gluon shadowing in Run~3~and~4 were evaluated in~\cite{Citron:2018lsq} in terms of the nuclear suppression factor $R_{\textup{Pb}}$ which is defined as the root square of the ratio of the photoproduction cross section $\sigma_{\gamma\textup{Pb}}$ measured in Pb–Pb UPC and the photoproduction cross section in the impulse approximation ($\sigma_{\rm IA}$) calculated as a reference photoproduction cross section off protons scaled by the integral over the squared Pb form factor~\cite{Guzey:2013xba}:
\begin{linenomath} 
\begin{equation*}
    R_{\textup{Pb}}(x) = \sqrt{\dfrac{\sigma_{\gamma\textup{Pb}}}{\sigma_{\textup{IA}}}},~~~~x = \dfrac{m_V}{\sqrt{s_{\textup{NN}}}} \textup{exp}(-y).
\end{equation*}
\end{linenomath}
Here, $m_V$ and $y$ are the mass and rapidity of the produced vector meson. Under the assumption that the coherent photoproduction cross section is proportional to the squared gluon density at the scale $Q = m_{V}/2$ this nuclear suppression factor can be used to constrain nuclear shadowing at different scales. The resulting pseudodata projections, based on EPS09~LO central values, are shown in Fig.~\ref{fig:rPb}.

Apart from heavy vector mesons measurements, there is also a possibility to study heavy flavor photoproduction, which is a direct probe of the gluon content of nuclei~\cite{Klein:2002wm}. The measurements of $\gamma + g \rightarrow c \Bar{c}~(b\Bar{b})$ are expected to be well feasible due to the high rates of these processes, despite low experimental acceptances and branching ratios \cite{Citron:2018lsq}. 

UPCs in ALICE may be possibly used to investigate light-by-light scattering process $\gamma\gamma \rightarrow \gamma\gamma$ and search for axion-like particle production $\gamma\gamma\rightarrow~a\rightarrow~\gamma\gamma$ in Run~3~and~4. The triggerless readout mode may allow ALICE to extend ATLAS and CMS measurements~\cite{Aad:2020cje,Sirunyan_2019,Citron:2018lsq} towards lower invariant masses.

\section{Event selection}

The upgrade of the ALICE experiment and the transition to the continuous readout mode will lead to a significant increase in data stream from the detector, therefore an advanced event selection is necessary in order to reduce the required storage volume.

In this contribution, several possible event selection strategies for central exclusive production and ultra-peripheral collision events are considered. The first strategy relies on online (high-level trigger) preselection of time frames containing CEP or UPC events without full event reconstruction and just using hardware information from forward detectors. The second strategy uses full real-time event reconstruction, that will be followed by asynchronous event selection. After the selection, the most relevant data will be recorded to the permanent storage. The third approach is based on recording of all the events to the permanent storage and full offline reconstruction followed by subsequent filtering of interesting events.

Measurements of CEP and UPC processes rely on the search for events with few tracks in the central barrel (e.\,g. two or four-prong events) in an otherwise empty detector~\cite{future_pp_programme}. The selection strategy for these events becomes particularly challenging in the continuous readout mode. The idea of the selection algorithm is based on using vetoes on the Fast Interaction Trigger detector (FIT) activity to reject non-diffractive (or non-UPC) events. However, in order to check activity in FIT for a CEP or UPC-like collision, one needs an accurate collision timestamp. The timestamp calculated using information from track matching in the Inner Tracking System (ITS) and the Time Projection Chamber (TPC) has a significant uncertainty of about 100~ns, with the typical time between interacting bunches being approximately equal to 25~ns in pp and 50~ns in Pb--Pb collisions. A more precise timestamp can be obtained if one of the tracks is matched to a hit in the Time-Of-Flight detector (TOF) therefore one has to search for events with at least one hit in TOF and zero activity in the FIT in the corresponding 25 ns time window.

\begin{figure}[h!]
\centering
\includegraphics[width=0.8\textwidth]{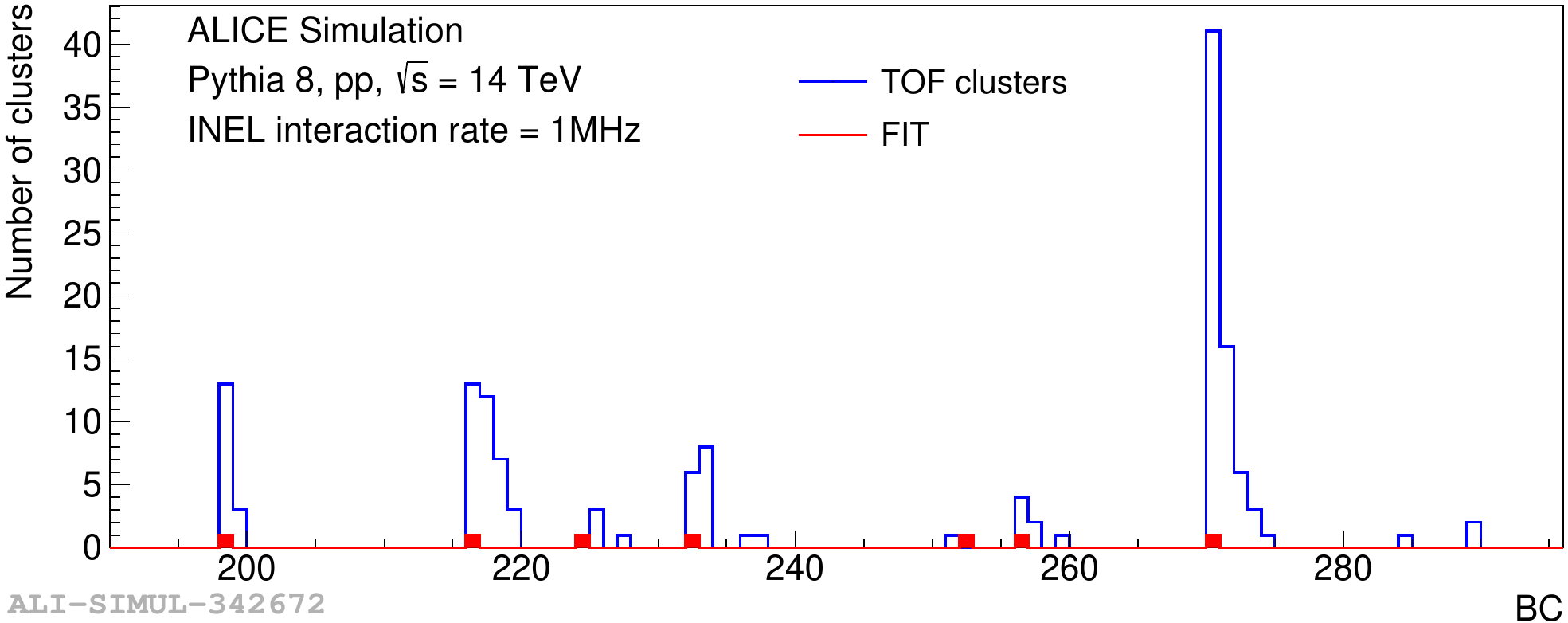}
\caption{The Pythia8 simulation: typical distribution of FIT and TOF clusters at 1~MHz interaction rate in pp at 14~TeV; <<BC>> or <<bunch crossing>> -- a typical time between interacting bunches~$\sim$~25~ns.}
\label{fig:fit_vs_tof}
\end{figure}

Feasibility studies on the online event preselection strategy were performed with the Pythia8 generator used to generate pp collisions at an interaction rate of 1~MHz at \break $\sqrt{s_{\textup{NN}}}=$ 14~TeV and the $\textup{O}^2$ framework used for the full ALICE detector simulation and event reconstruction. Figure~\ref{fig:fit_vs_tof} illustrates a typical time distribution of FIT and TOF clusters in one of the time frames. It can be seen that online preselection of time frames is hardly possible, since there are numerous bunch crossings (typical time between interacting bunches~$\sim$~25~ns) with zero FIT activity and only one or two clusters in TOF mainly originating from hadronic interactions in preceding bunch crossings. Consequently, FIT and TOF activity alone cannot be used for efficient time-frame rejection in the high-level trigger.

\begin{figure}[h!]
\centering
\includegraphics[width=0.75\textwidth]{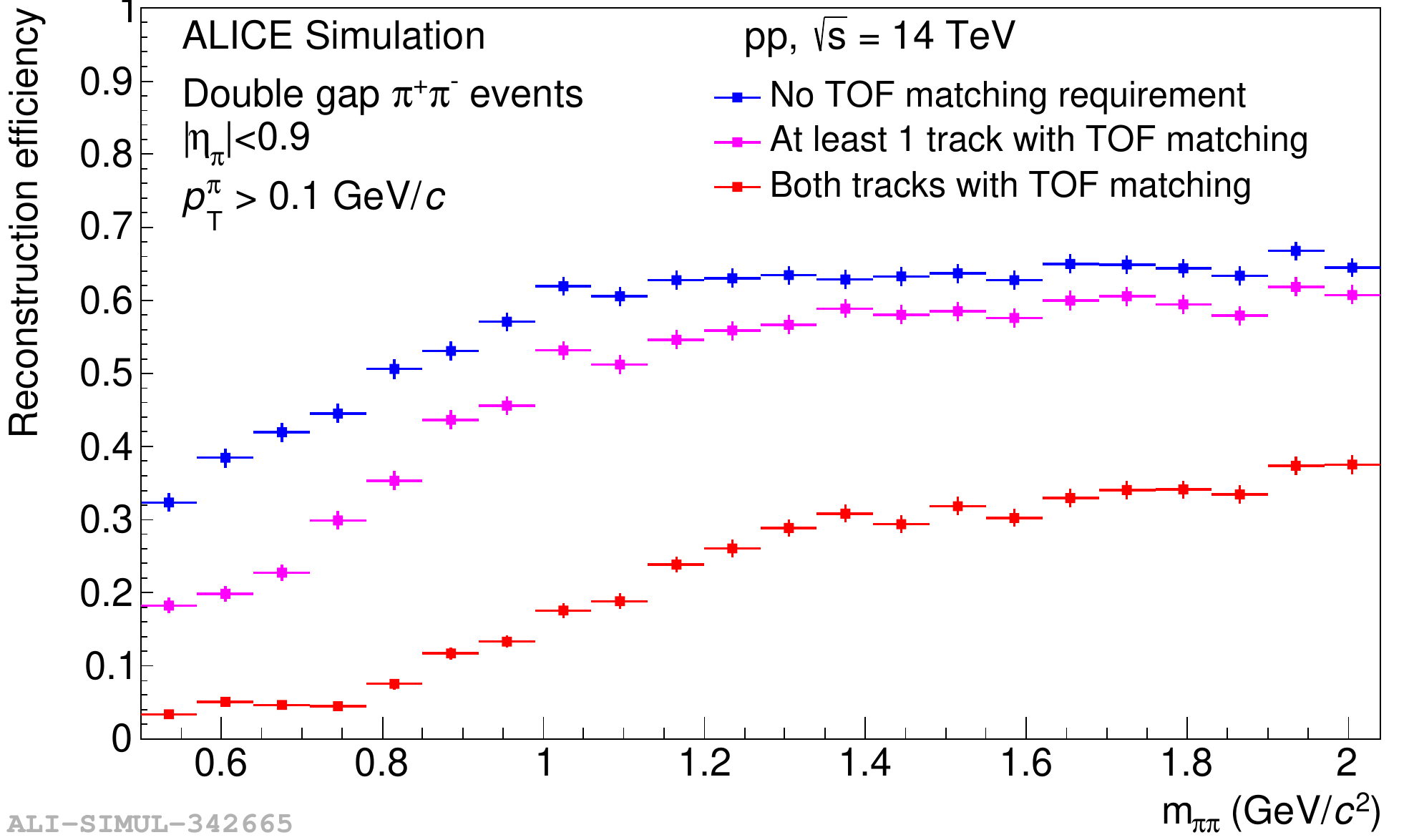}
\caption{The reconstruction efficiency for pion pairs in central diffractive events in pp at 14 TeV.}
\label{fig:reco_eff}
\end{figure}

Therefore, a strategy based on the full event reconstruction has been selected. The reconstruction efficiency of CEP-like events with two opposite-charge tracks and different TOF matching requirements was studied with a custom event generator. A dataset consisting of 90000 pion pairs was generated and the $\textup{O}^2$ framework was used for the full ALICE detector simulation in the continuous readout mode. Figure~\ref{fig:reco_eff} illustrates the reconstruction efficiency as a function of invariant mass  $m_{\pi\pi}$ of the pion pair for different TOF hit matching requirements (no TOF matching requirement, at least one track or both tracks with the TOF matching). There is a substantial loss of efficiency in the case of TOF matching requirement for both tracks, however there is no significant efficiency drop if only one TOF matched track is required. Since one TOF hit is enough for the matching of FIT and TOF information and FIT activity checks, a selection strategy based on the requirement of at least one track matched to TOF is considered as the most feasible option for CEP and UPC. 

\section{Conclusions}

In this contribution, the upgrade of the ALICE apparatus during the Long Shutdown~2 and the motivation for the studies of the central diffractive events in proton--proton collisions and ultra-peripheral heavy-ion collisions were discussed. The event selection strategies for CEP and UPC in Run~3 and Run~4 were considered and the feasibility studies on them were carried out. The results show that online event preselection using information just from FIT and TOF detectors is not possible, thus either full online reconstruction or offline reconstruction have to be done.

\bibliographystyle{pepan}
\bibliography{bibliography}

\end{document}